\documentclass[10pt, conference, letterpaper]{IEEEtran}
\usepackage[dvips,usenames]{color}
\usepackage{epsf}
\usepackage{times}
\usepackage{epsfig}
\usepackage{graphicx}
\usepackage{amsmath}
\usepackage{amssymb}
\usepackage{amsxtra}
\usepackage{here}
\usepackage{rawfonts}
\usepackage{times}
\usepackage{url}
\usepackage{cite}
\usepackage{multirow}

\twocolumn
\newtheorem{definition}{\bf Definition}

\newlength{\aligntop}
\newlength{\alignbot}
\makeatletter

\usepackage[normalem]{ulem}

\newcommand{\diag}{{\mathrm{diag}}}
\newcommand{\rank}{{\mathrm{rank}}}

\begin{document}

\title{\huge{Collaborative Spectrum Sensing from Sparse Observations Using Matrix Completion for Cognitive Radio Networks}}

\author{\authorblockN{ Jia (Jasmine) Meng$^1$, Wotao Yin$^2$, Husheng Li$^3$, Ekram Houssain$^4$ and Zhu Han$^1$}
\authorblockA{
$^1$ Department of Electrical and Computer Engineering, University of Houston\\}
$^2$ Department of Computational and Applied Mathematics, Rice University\\
$^3$ Department of Electrical Engineering and Computer Science, University of Tennessee at Knoxville\\
$^4$ Department of Electrical and Computer Engineering, University of Manitoba, Canada
}
\date{}

\maketitle
\begin{abstract}

In cognitive radio, spectrum sensing is a key component to detect spectrum holes (i.e., channels not used by any primary users). Collaborative spectrum sensing among the cognitive radio nodes is expected to improve the ability of checking complete spectrum usage states. Unfortunately, due to power limitation and channel fading, available channel sensing information is far from being sufficient to tell the unoccupied channels directly. Aiming at breaking this bottleneck, we apply recent matrix completion techniques to greatly reduce the sensing information needed. We formulate the collaborative sensing problem as a matrix completion subproblem and a joint-sparsity reconstruction subproblem.
Results of numerical simulations that validated the effectiveness and robustness of the proposed approach are presented. In particular, in noiseless cases, when number of primary user is small, exact detection was obtained with no more than 8\% of the complete sensing information, whilst as number of primary user increases, to achieve a detection rate of 95.55\%, the required information percentage was merely 16.8\%.

\end{abstract}

\section{Introduction}




Cognitive radio \cite{COR1,HBOOK} has been known as a novel paradigm for improving the utilization of the precious natural resource -- radio spectrum.
Spectrum sensing is a key component in cognitive radio for detecting spectrum holes, which are the spectrum channels not used by any primary user.
Since each cognitive radio (CR) node has only limited local observation to the whole spectrum due to various constraints,
collaborations among CR nodes are important for acquiring the complete spectrum information.

Related work \cite{GL05,ZZY05,GS05} introduces architecture and network protocols using either distributed or  centralized approaches for collaborative sensing in CR networks.
In the distributed approach, CR nodes exchange information using local common channels through distributed coordination. In the centralized approach,
a central control entity, e.g. a fusion center, gathers sensing information from all the CR nodes within a network through the common control channel.
In both cases, subject to energy constraints,  each CR user has very limited access to the whole spectrum. Furthermore, due to path loss, channel fading, and/or shadowing effects,  the  highly incomplete sensing information transmitted by CR users is affected by transmission loss or even error. Therefore, the challenge is for {\em the system to obtain complete channel states from their incomplete measurements}.


\textbf{Contributions:} 

We assume that a few of the $n$ channels are occupied by primary users and channel fading is not known. A system model is introduced in which each CR, instead of scanning each channel,  takes a small number of measurements that linearly combine multiple channels, which are sent to the fusion center. We allow some transmissions to fail so the fusion center receives a subset of the measurements. To assess the channel occupancy, we first argue that the matrix of all the measurements has a low-rank so it can be recovered by solving a matrix completion problem. Next, given the recovered matrix, the occupied channels and their fading values are reconstructed by solving a joint-sparsity reconstruction problem.
In our simulations we used FPCA, a matrix completion algorithm from \cite{MGC08}, and a novel joint-sparsity algorithm. In particular, in noiseless cases, when number of primary user is small, exact detection was obtained with no more than 8\% of the complete sensing information, whilst as number of primary user increases, to achieve a detection rate of 95.55\%, the required information percentage was merely 16.8\%.

This paper is organized as follows: In Section \ref{sec:model}, the system model is given. Algorithms for collaborative sensing, including the matrix completion and joint-sparsity parts, are described in Section \ref{sec:algorithm}. Simulation results are presented in Section \ref{sec:simulation}, and conclusions are drawn in Section \ref{sec:conclusion}.

\section{System Model}\label{sec:model}

Suppose there are $m$ CR nodes locally monitoring a subset of $n$ channels where $m<n$. A channel is either occupied by a primary user or unoccupied, corresponding to the states 1 and $0$, respectively. Assume that the number $s$ of occupied channels is much smaller than $n$. Our task is to tell the occupied channels from the CR nodes' observations.

\begin{figure}
  \center
  \includegraphics[width=2in]{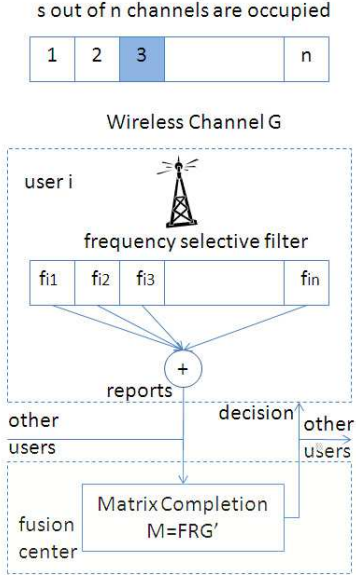}\\
  \caption{Collaborative Sensing using Matrix Completion}\label{f:system_model}
\end{figure}

The proposed approach for collaborative sensing is depicted in Figure \ref{f:system_model}. Instead of sensing one channel at a time, each CR takes measurements of multiple channels using the equipped frequency selective filters, and the measurements are sent to the fusion center. Supposing that totally $p$ reports regarding the $n$ channels are sent from a CR to the fusion center, we model this process by a $p\times n$ filter coefficient matrix $\mathbf{F}$. Let an $n \times n$ diagonal matrix $\mathbf{R}$ present the states of all the channels using 0 and 1 diagonal entries, indicating the unoccupied and occupied states, respectively. There are $s$ entries of 1 in $\diag(\mathbf{R})$. In addition, channel gains that affecting the CR nodes are described in an $m \times n$ channel gain matrix $\mathbf{G}$ given by
\begin{equation}\label{eqn:G}
\mathbf{G}_{i,j}=(d_{i,j})^{-\alpha/2}|h_{i,j}|,
\end{equation}
where $d_{i,j}$ is the distance from the primary transmitter using $j^{th}$ channel from the $i^{th}$ CR node, $\alpha$ is the propagation loss factor, and $h_{i,j}$ is the channel fading.
The measurement reports sent to the fusion center can be written as a $p\times m$ matrix
\begin{equation}
\mathbf{M}_{p \times m}=\mathbf{F}_{p \times n}\mathbf{R}_{n \times n}(\mathbf{G}_{m \times n})^\top.
\end{equation}

Measurement matrix $\mathbf{M}$ has the following two important properties \cite{CR08} required for completion from partial entries:
\begin{enumerate}
\item \textbf{Low Rank}: $\rank(M)$ equals to $s$, which is the number of prime users in the network and is usually very small.

\item \textbf{Incoherent Property}: Generate $\mathbf{F}$ randomly (subject to hardware limitation). From \eqref{eqn:G} and the fact that $\mathbf{R}$ has only $s$ nonzeros on the diagonal, $\mathbf{M}$'s SVD factors $\mathbf{U}$, $\Sigma$, and $\mathbf{V}$ satisfy the {\em incoherence condition} \cite{KO09}
\begin{itemize}
  \item There exists a constant $\mu_{0}> 0$ such that for all $i \in [p]$, $j \in [m]$, we have $\sum_{k=1}^{s} \mathbf{U}_{i,k}^{2} \leq \mu_{0} s$, $\sum_{k=1}^{s} \mathbf{V}_{i,k}^{2}$ $\leq \mu_{0} s$.
  \item There exists $\mu_{1}$ such that $\mid \sum_{k=1}^{s} \mathbf{U}_{i,k}\Sigma_{k} \mathbf{V}_{j,k} \mid \leq \mu_{1} s^{1/2}$.
\end{itemize}

\end{enumerate}

$\mathbf{M}$ is in general incomplete because of transmission failure. Moreover, each CR might only be able to collect a random (up to $p$) number of reports due to the hardware limitation.
Therefore, the fusion certain receives a subset set $\mathbf{E}\subseteq [p]\times[m]$ of $\mathbf{M}$'s entries. We assume that the received entries are uniformly distributed with high probability. Hence, we work with a model in which each entry shows up in $\mathbf{E}$ identically and independently with probability $\epsilon /\sqrt{p\times m}$. 
Given $\mathbf{E}_{p\times m}$, the partial observation of $\mathbf{M}$ is defined as a $p\times m$ matrix given by
\begin{equation}
{M}_{ij}^{E}=\left\{
\begin{array}{ll}
{M}_{ij}, & \mbox{if} \ (i,j)\in \mathbf{E},\\
0, & \mbox{otherwise}.
\end{array}
\right.
\end{equation}

We shall first recover the unobserved elements of $\mathbf{M}$ from $\mathbf M^E$. Then, we reconstruct $(\mathbf{R}\mathbf{G}^\top)$ from the given $\mathbf{F}$ and $\mathbf{M}$ using the fact that all but $s$ rows of $(\mathbf{R}\mathbf{G}^\top)$ are zero. These nonzero rows correspond to the occupied channels.
Since $p$ and $m$ are much smaller than $n$, our approach requires a much less amount of sensing and transmission, compared to traditional spectrum sensing in which each channel is monitored separatively.


\section{CR Sensing Matrix Completion Algorithm}\label{sec:algorithm}

In previous research on matrix completion \cite{CCS08, MGC08, KO09, DM09}, it has been proved that under some suitable conditions, a low-rank matrix can be recovered from a random, yet small subset of its entries by nuclear norm minimization:
\begin{equation}\label{mtx_cmp}
\min_{\mathbf{M}\in\mathbb{R}^{p\times n}} \tau\|\mathbf{M}\|_{\ast}+\frac{1}{2} \sum_{(i,j)\in\mathbf{E}}\left|\mathbf{M}_{i,j}-\mathbf{M}^{E}_{i,j}\right|^2,
\end{equation}
where $\|\mathbf{M}\|_\ast$ denotes the nuclear norm of matrix $\mathbf{M}$ and $\tau$ is a parameter discussed in Section \ref{sec:stop} below. For notational simplicity, we introduce the linear operator $\mathcal{P}$ that selects the components $\mathbf{E}$ out of a $p\times n$ matrix and form them into a vector such that $\|\mathcal{P}\mathbf{M}-\mathcal{P}\mathbf{M}^{E}\|^2_2=\sum_{(i,j)\in\mathbf{E}}| \mathbf{M}_{i,j}-\mathbf{M}^{E}_{i,j}|^2$. The adjoint of $\mathcal{P}$ is denoted by $\mathcal{P}^*$.



For our problem, we adopt FPCA by Ma et al. in \cite{MGC08}, which appears to run very well for the relatively small-dimensional application we focus on.
In the following subsections, we describe this algorithm and the steps we take for nuclear norm minimization. 
We further discuss the stopping criteria for iterations to acquire optimal recovery. Finally we show how to obtain $\mathbf{R}\mathbf{G}^\top$ from 
$\mathbf{M}$.

\subsection{Nuclear Norm Min. via Fixed Point Iterative Algorithm}

FPCA is based on the following fixed--point iteration
\begin{equation}\label{eqn:steps}
\left\{
\begin{array}{ll}
\mathbf{Y}^{k}=\mathbf{M}^{k}-\delta_{k}\mathcal{P}^*(\mathcal{P}\mathbf{M}^k-\mathcal{P}\mathbf{M}^{E}),\\
\mathbf{M}^{k+1}=S_{\tau\delta_{k}}(\mathbf{Y}^{k}),
\end{array}
\right.
\end{equation}
where $\delta_{k}$ is step size and $S_\alpha(\cdot)$ is the matrix shrinkage operator defined as follows:
\begin{definition}\label{def:shrinkage}
\textbf{Matrix Shrinkage Operator $S_{\alpha}(\cdot)$}:
Assume $\mathbf{M} \in \mathbb{R}^{p\times m}$ and its SVD is given by $\mathbf{M}=\mathbf{U}\diag(\sigma)\mathbf{V}^T$, where $\mathbf{U} \in \mathbb{R}^{p\times r}$, $\sigma \in\mathbb{R}_{+}^{r}$, and $\mathbf{V} \in \mathbb{R}^{m\times r}$. Given $\alpha>0$, $S_{\alpha}(\cdot)$ is defined as \begin{equation}
S_{\tau}(\mathbf{M}):= \mathbf{U}\diag\left(s_{\alpha}(\sigma)\right)\mathbf{V}^{T}
\end{equation}
with the vector $s_{\alpha}(\sigma)$ defined as:
\begin{equation}
s_{\alpha}(x):=\max\{x-\alpha,0\},~\mbox{component-wise.}
\end{equation}
\end{definition}

Simply speaking, $S_{\tau}(\mathbf{M})$ reduces every singular values (which is nonnegative) of $\mathbf{M}$ by $\tau$; if one is small than $\alpha$, it is reduced to zero. 

To understand \eqref{eqn:steps}, observe that the first step of \eqref{eqn:steps}  is a gradient-descent applied to the second term in \eqref{mtx_cmp} and thus reduce its value. Because the previous gradient-descent generally increases the nuclear norm, the second step of \eqref{eqn:steps} reduces the nuclear norm of $\mathbf{Y}^{k}$. Iterations based on \eqref{eqn:steps} converge when the step sizes $\delta_{k}$ are properly chosen (e.g., less than 2 or by line search) so that the first step of \eqref{eqn:steps} is not ``expansive'' (the other step is always non-expansive).

The second step of \eqref{eqn:steps} requires computing the SVD decomposition of $\mathbf{Y}^{k}$, which is the main computational cost of \eqref{eqn:steps}. However, if one can predetermine the rank of the matrix $\mathbf{M}$, or have the knowledge of the approximate range of its rank, a full SVD can simplified to computing only a rank-$r$ approximation to $\mathbf{Y}^{k}$. 
Specifically, the approximate SVD is computed by a fast Monte Carlo algorithm developed by Drineas et al.\cite{DKM06}. For a given matrix $\mathbf{A}\in\mathbb{R}^{m\times n}$ and parameters $k_s$, this algorithm returns an approximation to the largest $k_s$ singular values corresponding left singular vectors of the matrix $\mathbf{A}$ in a linear time.


\subsection{Stopping Criteria for Iterations}\label{sec:stop}

We tuned the parameters in FPCA for a better overall performance. Continuation is adopted by FPCA, which solves a sequence of instances of \eqref{mtx_cmp}, easy to difficult, corresponding to a sequence of varying (large to small) values of $\tau$. The final $\tau$ is the given one but solving the easier instances of \eqref{mtx_cmp} gives intermediate solutions that warm start the more difficult ones so that the entire solution time is reduced. Solving each instance of \eqref{mtx_cmp} requires proper stopping. 
We use the criterion:
\begin{equation}
\frac{\|\mathbf{M}^{k+1}-\mathbf{M}^{k}\|_{F}}{\max\{1,\|\mathbf{M}^{k}\|_{F}\}}<mtol
\end{equation}
where $mtol$ is a small positive scalar. Experiments shows that $1e^{-5}$ is good enough for obtaining correct detections.

\subsection{Channel Availability Estimation Based on the Complete Measurement Matrix}
Since $\mathbf{F}$ has more columns than rows, directly solving $\mathbf{X}:=\mathbf{R}\mathbf{G}^\top$ in \eqref{eqn:G} from given $\mathbf{M}$ is underdetermined. However, each row $X_i$ of $\mathbf{X}$ corresponds to the occupancy status of channel $i$. Ignoring noise in $\mathbf{M}$ for now, $X_i$ contains a positive entry if and only if channel $i$ is used. Hence, most rows of $\mathbf{X}$ are completely zero, so every column $X_{\cdot,j}$ of $\mathbf{X}$ is sparse and all $X_{\cdot,j}$'s are jointly sparse. Such sparsity allows us to reconstruct $\mathbf{X}$ from \eqref{eqn:G} and identify the occupied channels, which are the nonzero rows of $\mathbf{X}$.

Since the channel fading decays fast, the entries of $\mathbf{X}$ have a large dynamic range, which none of the existing algorithms can deal with well enough. Hence, we developed a novel joint-sparsity algorithm briefly described as follows. The algorithm is much faster than matrix completion and typically needs 1-5 iterations. At each iteration, every column $X_{\cdot,j}$ of $\mathbf{X}$ is independently reconstructed using the model $\min\{\sum_{i} w_i |X_{i,j}|: FX_{\cdot,j} = M_{\cdot,j}\}$, where $M_{\cdot,j}$ is the $j$th column of $\mathbf{M}$. For noisy $\mathbf{M}$, we instead use the constraint $\|FX_{\cdot,j} - M_{\cdot,j}\|\le \sigma$. The same set of weights $w_i$ are shared by all $j$ at each iteration. $w_i$ is set to 1 uniformly at iteration 1. After channel $i$ is detected in an iteration, $w_i$ is set to 0. Through $w_i$, joint sparsity information is passed to all $j$. Channel detection is performed on the reconstructed $X_{\cdot,j}$'s at each iteration. It is possible that some reconstructed $X_{\cdot,j}$ is wrong, so we let larger and sparser $X_{\cdot,j}$'s have more say. If there is a relatively large $X_{i,j}$ in a sparse $X_{\cdot,j}$, then $i$ is detected. We found this algorithm to be very reliable. The detection accuracy is determined by the accuracy of $\mathbf{M}$ provided.

\section{Simulation}\label{sec:simulation}

According to FCC and Defense Advance Research Projects Agency (DARPA) reports \cite{NTIA,FCC} data, we chose to test the proposed algorithm for spectrum utilization efficiency with such settings: at certain times, the number of active primary users varies from 1 to 3 on a given set of 100 channels with 20 CR nodes collaboratively detecting the occupied channels. The tests were performed at different sampling rates with different numbers of prime users (i.e., occupied channels). We define sampling rate as
\begin{equation}\nonumber
\frac{No.\ received\ measurements\  at\  the\  fusion\  center}{No.\ channels \times No.\ CRs},
\end{equation}
where $(No.\ channel \times No.\ CR)$ is the amount of total sensing workload in traditional spectrum sensing.
Performance was given in terms of probability of detection (POD) according to the definition in information-theory.
\begin{equation}\nonumber
POD=No.\ Hit/(No.\ Hit+No.\ Miss);
\end{equation}
where No. Hit is the number of successful detection of the appearance of  primary user(s), while No. Miss is the number of miss detections of the appearance of  primary user(s).

\begin{figure}
  \center
  \includegraphics[width=3in,height=2.5in]{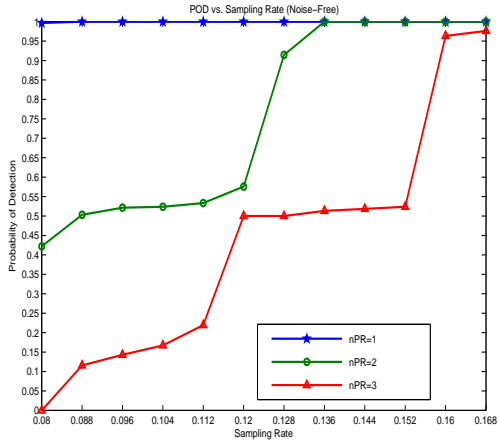}\\
  \caption{POD vs. Sampling Rate (Noise-Free)}\label{f:POD_SR_noisefree}
\end{figure}

\begin{figure}
  \center
  \includegraphics[width=3in,height=2.1in]{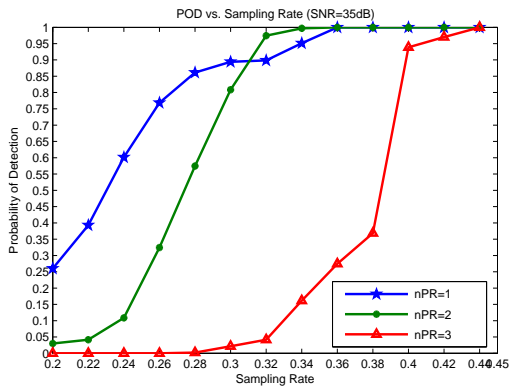}\\
  \caption{POD vs. Sampling Rate (SNR=35dB)}\label{f:POD_SR_noisy}
\end{figure}


Figure \ref{f:POD_SR_noisefree} shows the probability of detection at different sampling rates when there is no noise. In the relatively low spectrum usages cases, i.e. one out of 100 channels is occupied by a primary user, the proposed scheme enabled $100\%$ detection probability at a sampling rate as low as $8\%$. With 3 primary users appearing in the network, the detection task becomes harder. However, the algorithm still managed to realize a detection probability of higher than $95\%$ at a low sampling rate of $16.8\%$.

Figure \ref{f:POD_SR_noisy} shows the probability of detection at different sampling rates when the received signal is corrupted by Gaussian noise with a signal to noise ratio of $35dB$. Collaborative detection becomes even harder, we need higher sampling rate for exact primary user detection. Simulation result shows that, with $14.4\%$ sampling rate, 1 or 2 primary user(s) can be detected exactly. As the number of primary users increases, more samples are needed.

As we can see from the simulation results, the proposed approach achieves very high probability of detection at an extremely low sampling rate compared to traditional spectrum sensing. We plan to improve our algorithms and perform broader experiments. The results will be reported in a forthcoming journal paper.

\section{Conclusions}\label{sec:conclusion}
In this paper, we propose a collaborative spectrum sensing approach to detect spectrum holes in a cognitive radio network. We model the collaborative detection problem as a matrix completion problem in which partial observations of a matrix enable its faithful reconstruction. We solve the matrix completion problem by the recent algorithm FPCA and estimate channel availability based on joint sparsity recovery. Performance of the proposed approach was tested for frequency utilization efficiency ranging from 1\% to 3\%. In the noiseless cases, exact detection was obtained with no more than 8\% of the complete sensing information, whilst as the number of primary user increases, to achieve a detection rate of 95.55\%, the required information percentage was merely 16.8\%. In the noisy cases (SNR: 35dB), less than $15\%$ samples enabled exact detection of small numbers of primary  users. To summarize, the proposed approach significantly reduces the amount of sensing and transmission workload of cognitive radios for wide range spectrum sensing.


\section*{Acknowledgements}
The work of W. Yin was supported in part by NSF CAREER Award DMS-07-48839, ONR Grant N00014-08-1-1101, and an Alfred P. Sloan Research Fellowship.
The work of H. Li was supported in part by NSF 0831451 and NSF 0901425.
The work of Z. Han was supported in part by NSF CNS-0910461, CNS-0901425, and Air Force O¡Àce of Scientific Research.

\bibliographystyle{IEEEbib}

\end{document}